\documentstyle[sprocl]{article}
\def\be{\begin{equation}}
\def\ee{\end{equation}}
\def\bea{\begin{eqnarray}}
\def\eea{\end{eqnarray}}

\begin{document}

\title{A VARIETY OF GIANT RESONANCES IN METAL CLUSTERS}

\author{V.O. NESTERENKO$^{1,2}$, W. KLEINIG$^{2,3}$ and F.F. de SOUZA CRUZ$^{1}$}

\address{$^{1}$Departamento de Fisica, Universidade Federal de Santa 
Catarina,\\ Florian$\acute o$polis, SC, 840000-900, Brasil\\
E-mail: nester@fsc.ufsc.br, fred@fsc.ufsc.br}

\address{$^{2}$Bogoliubov Laboratory of Theoretical Physics,
Joint Institute for Nuclear Research, 
Dubna, Moscow region, 141980, Russia\\ E-mail: kleinig@thsun1.jinr.ru}

\address{$^{3}$Technische Univerrsit$\ddot a$t Dresden, 
Institut f$\ddot u$r Analysis, Dresden, D-01062, Germany}

\maketitle

\abstracts{\  Multipole electric and magnetic giant resonances in metal
clusters are reviewed and compared with their counterparts in atomic nuclei.
The main attention is paid to E1 resonance (dipole plasmon).}

\section{Introduction}

Metal cluster (MC) is a bound system consisting of atoms of some metal. The
amount of atoms can vary from a few to many thousands. Some MC, mainly of
alkali (Li, K, Na, ...) and noble (Ag, Au, ...) metals, demonstrate a
striking similarity to atomic nuclei (see reviews$^{1-5}$). In these
clusters the valence electrons are {\it weakly} coupled to the ions and,
like nucleons in nuclei, are not strongly localized. The mean free path of
valence electrons is of the same order of magnitude as the size of the
cluster. This favors the valence electrons to form a mean field of the same
kind as in nuclei (with the similar shell structure and magic numbers). In
addition to the mean field, MC demonstrate other similarities with atomic
nuclei: deformation in the case of open shells, variety of giant resonances
(GR), fission, etc.. As a result, many theoretical ideas and methods of
nuclear physics can, after a certain modification, be applied to MC \cite
{Ne92,Br93,Ne_Dub}.

This review is devoted to collective oscillations of valence electrons in
MC. Valence electrons can be considered as the counterparts of nucleons in
nuclei, and their oscillations as the counterparts of nuclear GR.
Investigation of GR in MC is interesting in two aspects: it allows to
understand deeper general properties of collective modes in finite Fermi
systems and, simultaneously, allows to study striking peculiarities of MC.
GR in clusters and atomic nuclei are well overlapped. However,  some
specific properties of MC cause considerable differences in the behavior of
GR in these two systems. 
For example: the Coulomb interaction and the ''spill-out'' effect provide a
specific dependence of GR properties on the mass number; the negligible
character of the spin-orbital interaction leads to the decoupling of spin
and orbital magnetic modes; clusters can have much more particles (atoms)
than nuclei, which favors very strong orbital magnetic resonances; for most
of the clusters the role of the ionic subsystem is important; at different
temperatures MC can be in solid, liquid and even ''boiling'' phases, which
greatly influences GR properties; characteristics of GR vary considerably
whether the clusters are charged or neutral, free or embedded to a
substrate, pure or with impurities atoms, etc..

Our consideration will be limited by certain physical conditions.

-- i) The modern techniques allow to fabricate atomic clusters from atoms of
about any element of the periodic table. However, the conception of the mean
field for valence electrons is realized only for a minority, -- mainly for
clusters of alkali and noble metals and, in a less extent, for neighboring
elements. So, we should limit ourselves by this MC region.

-- ii) In some alkali metals (Na and K) the ionic lattice can, to good
accuracy, be replaced by a uniform distribution of the positive charge over
cluster's volume. This is so-called jellium approximation which greatly
simplifies the analysis and calculations. This approximation is enough for
the description of many properties of alkali MC and will widely be used in
the review. However, it often fails beyond Na and K and then a more explicit
treatment of the ionic structure is necessary \cite{Br93,Bre_94rev}.

-- iii) The ionic subsystem is supposed to be ''frozen''. 
In the review only collective oscillations of valence electrons will be
considered. 

-- iv) The validity of jellium approximation is supported by temperature
fluctuations of ions, which smooth ion positions. It fails in the low
temperature region (approximately at $T<100$ K) where the explicit treatment
of the ionic structure is important. At too high temperatures ($T>1000$ K)
the quantum shells of the mean field are washed out, what establishes an
upper limit for our considerations. We will consider GR in a temperature
interval between these extreme cases.

\section{Theoretical Grounds}

\label{sec:the}

Due to the similarity between MC and nuclei, many models of nuclear theory
have been applied to study MC\cite{Ne92,Br93}. In due time, some of them
have been introduced to nuclear physics from solid body field and then
subsequently modified to describe {\it finite} Fermi systems. Now they turn
out to be useful for clusters. In particular, a large variety of the RPA
methods have been adopted , scaling from simple versions, like the sum rule
approach$^{6-9}$ 
and the local RPA \cite{RB90,ReGB_AP96}, to sophisticated full RPA models,
like time-dependent Hartree-Fock \cite{GJ92} and time-dependent local
density approximation (TD-LDA)$^{13-26}$ (for a more complete list of
citation see Refs.\cite{Ne92,Br93,Rubrev}). The simple models can describe
the gross structure of GR but not the fragmentation of the collective
strength. The full RPA models can describe the fragmentation but are very
time consuming. The last shortcoming becomes crucial for deformed and large
spherical clusters where the number of particles, and thus the size of the
configuration space, is very large. In this connection, the intermediate
class of the models, the RPA with {\it separable} residual forces (SRPA),
seems to be very promising$^{7,27-35}$. 
The separable ansatz allows one to turn the RPA matrix into a simple
dispersion relation. This drastically simplifies the eigenvalue problem
preserving, at the same time, the main advantage of the full RPA to describe
the fragmentation of the collective strength. The SRPA version derived in
Refs.$^{27-33}$ provides the accuracy of full RPA calculations \cite
{KNRS_EPJ98}, can be applied to systems of any shape \cite
{Ne_ZPD,Ne_PRC,NK_Tsu,NKK_Pra}, and allows to treat GR in MC and atomic
nuclei on the same microscopic footing \cite{NK_PS,Ne_PRC,NKK_Pra}. The
results obtained within this SRPA version will be widely used in the review
as illustrative examples.

For the description of collective oscillations, the SRPA, and most of the
other models, exploit, as a starting point, the Kohn-Sham energy functional 
\cite{KS65,GL76} for a system of $N_e$ valence electrons:

\[
E\{n({\bf r},t),m({\bf r},t),\tau ({\bf r},t)\} =1/2\int \tau ({\bf r},t)d%
{\bf r}+\int v_{xc}(n({\bf r},t),m({\bf r},t))d{\bf r} 
\]
\begin{equation}
+1/2\int \int \frac{(n({\bf r},t)-n_i({\bf r}))(n({\bf r_1},t)-n_i({\bf r_1}%
))}{|{\bf r}-{\bf r_1}|}d{\bf r}d{\bf r_1},
\end{equation}
which includes the kinetic energy, the exchange-correlation term in the
local density approximation (LDA) \cite{GL76,VWN80} and the Coulomb
interaction, respectively. Here, $n({\bf r},t)=n({\bf r},t)\uparrow +n({\bf r%
},t)\downarrow =\sum\nolimits_l|{\phi }_l({\bf r},t)|^2$, $m({\bf r},t)=n(%
{\bf r},t)\uparrow -n({\bf r},t)\downarrow $ and $\tau ({\bf r}%
,t)=\sum\nolimits_l|\bigtriangledown {\phi }_l({\bf r},t)|^2$ are the
density, magnetization density (z-component) and kinetic energy density of
valence electrons, respectively; $n_i({\bf r})$ is the ionic density in the
jellium approximation; $\phi _l({\bf r},t)$ is a single-particle wave
function. The convention $e=m_e=\hbar =1$ is used. The functional (1) can
have additional terms if the ionic structure is treated beyond the jellium
approximation.

The time-dependent single-particle Hamiltonian is obtained as 
\begin{equation}
H({\bf r},t)\phi _l({\bf r},t)=\frac{\delta E}{\delta \phi _l^{*}({\bf r},t)}%
.
\end{equation}
In the small-amplitude limit of a collective motion, the densities can be
written as $n({\bf r},t)=n_{{0}}({\bf r})+\delta n({\bf r},t)$ and $m({\bf r}%
,t)=m_{{0}}({\bf r})+\delta m({\bf r},t)$ where $n_0({\bf r})$ and $m_0({\bf %
r})$ are the static ground state densities ($m_0({\bf r})=0$ in spherical
and unpolarized clusters) and the values $\delta n({\bf r},t)$ and $\delta m(%
{\bf r},t)$ are small time-dependent density variations (transition
densities). Then, in the linear approximation to the density variations, the
Hamiltonian (2) is a sum of the static and dynamical parts. The static part 
\begin{equation}
H_0({\bf r})=T+V_0({\bf r})=-\frac{\triangle }2+(\frac{dv_{xc}}{dn})_{n=n_{{0%
}},m=m_{{0}}}+\int \frac{n_{{0}}({\bf r_1})-n_i({\bf r_1})}{|{\bf r}-{\bf r_1%
}|}d{\bf r_1}
\end{equation}
constitutes the Kohn-Sham single-particle potential. It can be approximated
with a good accuracy by phenomenological potentials, such as the harmonic
oscillator\cite{LiS89_ZPD} (for small spherical MC), Nillson-Clemenger\cite
{C85,RFB95} (for deformed MC) or Woods-Saxon \cite{Ne_PRA,NHM90,FP96} (for
spherical and deformed MC .

In the electric channel, the dynamic part of the Hamiltonian (residual
interaction) has the form 
\begin{equation}
\delta H({\bf r},t)=(\frac{d^2v_{xc}}{dn^2})_{n=n_0}\delta n({\bf r},t)+\int 
\frac{\delta n({\bf r_1},t)}{|{\bf r}-{\bf r_1}|}d{\bf r_1}.
\label{eq:ELres}
\end{equation}
The dominant term here is the Coulomb interaction. The residual interaction
in this channel is always positive (repulsive) and shifts the unperturbed
electrical multipole strength from the typical particle-hole (ph) values $%
\omega _{ph}=0.9-1.5$ eV to higher energies 2.6-3.2 eV.

In the spin channel, the dynamical part is 
\begin{equation}
\delta H({\bf r},t)=(\frac{d^2v_{xc}}{dndm})_{n=n_0,m=m_0}\delta n({\bf r}%
,t)\delta m({\bf r},t).  \label{eq:MLres}
\end{equation}
Here, the residual interaction is determined by the exchange-correlation
term. It is always negative (attractive) and shifts the unperturbed magnetic
multipole strength from $\omega _{ph}=0.9-1.5$ eV to lower energies 0.2-0.8
eV. The residual interactions (4) and (5) are typical for the TD-LDA
calculation scheme.

In what follows we will mainly consider clusters constituted from monovalent
atoms, like alkali metals, for which the numbers of valence electrons and
atoms coincide, $N_e=N$.

\section{Electric Dipole Giant Resonance (E1 GR)}

Unlike nuclei, where different kinds of GR are well investigated both
experimentally and theoretically, our knowledge in clusters is mainly
limited by the electric dipole resonance (dipole plasmon). Experimentally
the E1 GR has been observed in a variety of clusters: small and large,
spherical and deformed, neutral and charged, hot and cooled (see, for
example, Refs.$^{43-49}$). 
As a rule, the photoabsorption cross section was measured by methods of the
depletion spectroscopy. For other GR ($EL(L\ne 1)$, $ML$) there are only
theoretical predictions$^{6-9,28,30,32,50,51}$. 

Physical interpretations of E1 GR in clusters and nuclei are very similar:
while in nuclei it is caused by translations of neutrons against and
protons, then in clusters it is a result of translations of the valence
electron against ions \cite{YanB_PR91}. In spite of this similarity, the
dipole resonance in clusters exhibits many interesting peculiarities which
will be discussed below.

\subsection{Energy of E1 GR: Step by Step}

The description of E1 energy for clusters is a rather complicated task. For
example, while in nuclei its energy depends on the mass number as $A^{-1/3}$%
, in clusters the E1-energy can both decrease (Ag clusters) and increase
(alkali MC) with the number of atoms. Let us consider this important
characteristic step by step.

{\bf Step one: Mie frequency and spill-out effect}. In the simplest
approximation, MC can be considered as a {\it classical} metallic drop.
Then, the E1-energy is described by Mie expression \cite{Mie}: $\omega
_{Mie}=\omega _p/\sqrt{3}$ where $\omega _p$ is the plasma frequency. For Na
clusters $\omega _{Mie}=3.41eV$. This value is much higher than the
experimental E1-energy which is 2.5-2.8 eV for spherical Na clusters with $%
N<100$.

The agreement with the experiment is considerably improved if we take into
account the {\it quantum} spill-out effect. This effect means that since the
valence electrons are quantum entities, they are not well localized and so,
unlike the classical ionic jellium, 
can be partly {\it spilled out } beyond the jellium boundary. In principle,
this effect takes place in any two-component quantum system including atomic
nuclei and atoms (a ''neutron skin'' in small nuclei is a relevant example).
With the spill-out, the E1 energy in MC is described as \cite{LiS89_ZPD} 
\begin{equation}
\omega _{E1}=\omega _{Mie}(1-\frac 12\frac{\delta N_e}{N_e})
\end{equation}
where $\delta N_e$ is the number of spilled out valence electrons. As a
result, the discrepancy with the experiment reduces to 0.2-0.3 eV. The
spill-out effect allows to explain the increase of the E1-energy with N,
observed in alkali MC. The value $\delta N_e$ decreases with the size (for
example, $\delta N_e=1.5(19\%)$ and $9.5(7\%)$in $Na_8$ and $Na_{138}$,
respectively \cite{Ar_NC89}) leading to the corresponding increase in the
E1-energy.

{\bf Step two: beyond jellium approximation, ionic structure, local and
nonlocal effects.} The remaining discrepancy can be removed in a large
extent by the explicit treatment of the ionic subsystem. First of all, we
should take into account that ions are not the points but have a size.
Inside this size, ion core electrons (ICE) (do not confuse them with the
valence electrons) screen the pure Coulomb interaction of ions and valence
electrons. To take into account this screening, the atomic pseudopotentials
(PP) are used (see, for instance, \cite{BG95,CCG95,YB96,BHS82}). They allow
to describe correctly the spectrum of valence electrons in isolated atoms 
{\it without} the solution of the complicated many-body atomic task. Being a
sum of contributions of ICE with different orbital momenta, PP have {\it %
local} (s-electrons) and {\it nonlocal} (p and d electrons) parts \cite
{BHS82}. To avoid dealing with nonlocal functions, the Pseudo-Hamiltonians
(PH) were introduced as the next simplifying step \cite{BCC89,L96}. PH,
being derived from PP, lead to less involved (but with the same accuracy)
calculations since, unlike the PP, they treat the nonlocality only through
the differential operators. Folding atomic PH with jellium, one gets PH for 
{\it atomic clusters}$^{19-21}$. PH have the additional advantage to be
easily incorporated to the common calculation schemes.

As compared to the conventional Kohn-Sham Hamiltonian, PH include the
additional local and two non-local (the orbital contribution and the
effective mass) terms. As is seen from Figure~1, in K
clusters (the same for Na) the nonlocal contributions are negligible and the
local term is enough to get good description of the E1-energy\cite
{KNRS_EPJ98}. This is not the case for Li clusters, where only both, local
and nonlocal, contributions provide the agreement with the experiment \cite
{KNRS_EPJ98}. In some studies (see, e.g., Ref.~\cite{BG95}) the ICE effects
are taken into account together with some averaged treatment of the ionic
arrays in a cluster. The latter leads to an additional, but rather moderate,
redshift of the E1-energy.

{\bf Step three: direct dynamical ICE contribution.} The ICE effects
discussed above are realized through the change of the single-particle
characteristics with the subsequent 
renormalization of the residual interaction. Besides this {\it indirect}
way, the ICE can {\it directly} influence the dynamics and thus lead to new
peculiarities of the E1 GR. This can be well demonstrated for Ag clusters
where, like atomic nuclei and {\it unlike} alkali MC, the E1-energy {\it %
decreases} with a size \cite{Ti_92GR}. The physics behind is that in these
clusters the energy of ICE excitations is comparable to the E1-energy. The
coupling of these two modes additionally screens the interaction of valence
electrons with ions (direct dynamical ICE contribution) and finally causes
the redshift (decrease) of the E1-energy \cite{SR97}. This effect is mainly
of a volume character and so intensifies with a cluster size. As a result,
the E1 energy in Ag clusters decreases with N. This tendency overpowers the
opposite one caused by the spill-out effect.

In alkali clusters, where the ICE excitations have energies much higher than
the E1 GR, the direct dynamical ICE contribution can be neglected and the
evolution of the E1-energy with the cluster size is determined mainly by the
spill-out effect.

\section{Landau Damping and Width of E1 GR}

The main physical mechanisms forming the plasmon width are the thermal
fluctuations of a cluster shape and the Landau damping (RPA fragmentation of
the collective strength) \cite
{Br93,YVB93,KNRS_EPJ98,R_SRPA,YB_APNY91,BL95,MonRei_temp}. The relative
contributions of these two mechanisms change with a cluster size. As is seen
from Figure~2, in small clusters, like $Na_{21}^{+}$, where
the Landau damping is negligible, the thermal fluctuations determine about
all the width. In clusters of moderate size, like $Na_{59}^{+}$, the Landau
damping is stronger and greatly contributes to the width. This is especially
the case for deformed clusters. In large clusters, like $Na_{441}^{+}$, the
Landau damping is weaker though its contribution to the width remains to be
considerable.

Figure~2 (bottom) shows that the Landau damping is closely
related with the shell structure \cite{KNRS_EPJ98}. In $Na_{21}^{+}$ the
dipole plasmon lies in the wide gap between the bunches of $\Delta {\cal N}$%
=1 and $\Delta {\cal N}$=3 particle-hole (ph) states and remains almost
unperturbed as a collective peak. With increasing the cluster size, the
resonance approaches the bunch $\Delta {\cal N}$=3 and, in $Na_{59}^{+}$, is
already interferes with ph states of this bunch, which leads to the
considerable Landau damping. For larger clusters, the plasmon runs to the
swamp of ph states. This leads to a general trend of increasing the width
which is, however, overlaid by sizeable fluctuations \cite{KNRS_EPJ98,R_SRPA}%
. But here a further mechanism comes into play: the coupling between the
resonance and ph states fades away due to increasing mismatch of $\Delta 
{\cal N}$=1 ph configurations (which mainly generate the plasmon) and
surrounding ph states with much larger values of $\Delta {\cal N}$. This
finally leads to a decrease of the plasmon width $\propto N_e^{-1/3}$
estimated analytically in the wall formula \cite{YB_APNY91} and tested in
the RPA calculations \cite{R_SRPA}.

The Landau damping in MC with $N<40$ is rather sensitive to cluster charge:
being strongest in negatively charged ones (anions), the Landau damping is
considerably reduced while passing to neutral and then to positively charged
clusters (cations) \cite{Yan_CPL}. This effect is caused by a strong
dependence of the single-particle potential depth $V_0$ on the cluster
charge. In anions the potential is shallow ($V_0\simeq -2$ eV), the energy
gaps between $\Delta {\cal N}$ bunches are very smooth and, so, there are
good conditions for a sizeable Landau damping (see discussion above). In
neutral clusters and more in cations, the potential depth is increased to
about -7 eV, the gaps between $\Delta {\cal N}$ bunches in the ph spectrum
become more distinctive, which weakens the Landau damping.

\section{Temperature Effects}

In most of experiments with GR in MC, the typical cluster temperature is
estimated to be in the interval 300-900 K which corresponds to the thermal
energy $kT=0.03-0.09$ eV. At these temperatures, ions behave as classical
particles and quantum properties of the cluster are mainly determined by
valence electrons. This can be easily proved \cite{B90} by using the
uncertainty relation $\Delta x\Delta p\geq \hbar $. This relation gives
lower bounds for the momentum and energy of a particle in a system: $\Delta
p=\hbar /\Delta x$ and $\Delta E=(\Delta p)^2/2m$, respectively. Taking $%
\Delta x\leq 1.5\AA $ (the diameter of $Na_{20}$)  for both ions and valence
electrons, one gets 
\[
\Delta E_e\geq 0.16eV\qquad \mbox{for electrons,} 
\]
\[
\Delta E_i\geq 10^{-4}eV\quad \mbox{for ions.} 
\]
The energy of a quantum motion of valence electrons considerably exceeds the
thermal energy, which favors their quantum behavior. The opposite situation
takes place for ions which, therefore, should exhibit the classical
behavior. Such result takes place because the ionic mass is much larger than
the electron one.

The difference in ionic and electron masses leads to other interesting
consequence. Namely, almost all the thermal energy is contained in the ionic
subsystem. Valence electrons are embedded to the thermal ionic bath. So,
unlike atomic nuclei, MC represent the case of the canonical ensemble.

The bulk melting points for K, Na and Li are $T_b=336$, 371 and 452 K,
respectively. This means that most of the measurements for GR in MC have
been done for clusters in a liquid-like phase.

As was mentioned above, in small clusters, thermal shape fluctuations
provide the dominate contribution to the plasmon width. It is interesting
that, while in nuclei these fluctuations are mainly of a quadrupole form, in
MC they are mainly octupole \cite{MonRei_temp}. The reason is that spherical
and neighboring MC are soft to the octupole deformation.

Photoexcitation is a rapid process in the ionic time scale. So, every
response of a cluster represents its instantaneous shape and the
experimental cross section gives a properly weighted response of all allowed
shapes \cite{BL95}.

The higher the temperature, the larger the plasmon width and the smaller the
plasmon energy. The temperature shift is estimated as about $1\%$ of the
plasmon energy per 100 K \cite{Bre91,HB67}. It can be explained by the
effectively increase of the cluster size with temperature. The larger the
size, the bigger the static dipole polarizability which is expressed through
the cluster radius as $\alpha _{E1}=R^3$. The polarizability is connected
with the plasmon energy through the inverse sum rule, $\alpha
_{E1}=2m_{-1}\simeq <E1>^2/\omega _{E1}$. So, the higher the temperature,
the larger $\alpha _{E1}$ and, consequently, the smaller $\omega _{E1}$.

Recent experiments show that at sufficiently low temperatures the
gross-structure of the E1 GR drastically changes \cite{Na10_PRL}. For
example, the axially deformed cluster $Na_{11}^{+}$ at 380 K demonstrates
the typical two-peak spectrum determined by the deformation splitting of E1
GR. At 35 K the same resonance exhibits much more complicated structure
including at least 6 well-distinguished peaks. This structure reflects the
ionic arrangement which is not washed out at so low temperature by
variations of ions. In this case, the jellium approximation is not valid and
models based on this cannot be applied. The E1 GR in small clusters at low
temperature seems to be best described by {\it ab initio} quantum-chemical
calculations \cite{BK_Pra}.

\section{E1 GR in Deformed Clusters}

Like nuclei, MC with open shells have quadrupole deformation$^{45-49,64-68}$%
. 
There are experimental indications of both prolate and oblate axial
quadrupole shapes, as well as of $\gamma $-deformation$^{45-49}$. 
In the framework of different methods (Strutinski's shell correction method,
ultimate jellium model, etc.) hexadecapole and octupole deformations as well
as high isomerism have been predicted$^{64-68}$. Rather strong quadrupole,
hexadecapole and octupole deformations should take place at least up to MC
with $N\sim 700$ \cite{FP96Er}. 
Like in nuclei, E1 GR in axially deformed MC exhibits the deformation
splitting in two peaks (see Figure~3). The right peak is about
twice larger than the left one in prolate clusters (see $Na_{11}^{+}$, $%
Na_{15}^{+}$, $Na_{27}^{+}$) and, vice versa, in oblate clusters (see $%
Na_{35}^{+}$).

Most of MC are deformed. But getting an experimental information on a
cluster shape, even in the simplest case of a quadrupole deformation, is
rather nontrivial problem. In nuclei rotational bands serve as source of
such information. In principle, deformed clusters can rotate. But, due to a
large value of the moment of inertia, rotational energies are very small
and, being of the same order of magnitude as the thermal energy, fail to be
observed. In this connection, the splitting of E1 GR in deformed clusters is
now a {\it single direct} manifestation of quadrupole deformation and the
main source of the information about it.

\section{Multipole GR, Asymptotic Trends, Restoring Forces}

So far, the depletion spectroscopy methods (photoabsorption and
photofragmentation) were mainly exploited for observation of E1 GR in MC 
\cite{He93}. The other reactions $((e,e^{\prime }),(\gamma ,\gamma ^{\prime
})$ and etc.) are not yet sufficiently developed, which impedes the
observation of other GR. The similar situation took place in nuclear physics
in early seventies. For this reason an investigation of EL GR with $L\ne 1$
is yet limited to theoretical predictions$^{6-9,28,30,32,50,51}$. In Figure 
4, E2 and E3 GR in spherical $Na_{59}^{+}$, calculated within
the SRPA, are presented as typical examples.

It is instructive to consider the main trends of electrical multipole giant
resonance with the size (N) and multipolarity (L), and also the origin of
the GR restoring forces. Such analysis has been done within the sum rule
approach (SRA) in Ref.\cite{Se_SRA_PRB89}. In the jellium approximation for
valence electrons, $n_{{0}}(r)=n_i(r)=n^{+}\theta (r-R)$ (the spill-out
effect is neglected), the energy of $EL(L\ne 0)$ GR can be written as \cite
{Se_SRA_PRB89} 
\begin{equation}
\omega _{EL}=\sqrt{\frac{m_3}{m_1}}=\hbar \sqrt{\frac 23(2L+1)(L-1)\frac{%
\beta _F^2}{R^2}+\omega _p^2\frac L{2L+1}}  \label{eq:omEL}
\end{equation}
where $m_1=\sum_iB(EL,gr\to i)\omega _i$ and $m_3=\sum_iB(EL,gr\to i)\omega
_i^3$ are the sum rules, $\beta _F=(3/5)^{1/2}(3\pi ^2)^{1/3}\frac
1mn_0^{1/3}$ and R is the radius of a cluster. The first term in Eq. \ref
{eq:omEL} is the contribution of the kinetic energy (the similar expression
have been obtained earlier in Ref. \cite{NS80}). The second term is
determined by the Coulomb interaction between valence electrons (ee) and
valence electrons and ions (ei). Eq. \ref{eq:omEL} shows that E1 GR is
determined only by the Coulomb interaction. 
In the limit of large R, one has 
\begin{equation}
\omega _{EL}\to \hbar \omega _p\sqrt{\frac L{2L+1}}.
\end{equation}
The larger $L$, the higher the excitation energy of the GR. In general, due
to the first term in Eq. \ref{eq:omEL}, the energy of $EL(L\ne 0,1)$ GR is
decreased with N. For low L in small clusters this tendency is changed by
the spill-out effect. 

The separate analysis for E0 GR gives the increase of the E0 energy with N
to the limit $\omega _{E0}\to \hbar \omega _p$.

\begin{table}[tbp]
\caption{Relative contributions to $m_3$ for $Na_{92}$: kinetic energy $%
(m_3(T))$, exchange and correlations $(m_3(xc))$, electron-electron
interaction $(m_3(ee)$, electron-ion interaction $(m_3(ei))$ and total
Coulomb interaction $(m_3(C)=m_3(ee)+m_3(ei))$ \protect\cite{Se_SRA_PRB89}. }
\label{tab:m3EL}\vspace{0.4cm}
\par
\begin{center}
{\footnotesize 
\begin{tabular}{|c|c|c|c|c|c|}
\hline
L & $m_3(T)$ & $m_3(xc)$ & $m_3(ee)$ & $m_3(ei)$ & $m_3(C)$ \\ \hline
1 & 0 & 0 & 0 & 1 & 1 \\ 
2 & 0.08 & 0 & -0.77 & 1.69 & 0.92 \\ 
5 & 0.51 & 0 & -2.29 & 2.78 & 0.49 \\ \hline
\end{tabular}
}
\end{center}
\end{table}

It is seen from Eq. \ref{eq:omEL} that the value $m_3$ has the meaning of a
restoring force \cite{RB90}. In Table \ref{tab:m3EL} the contributions to $%
m_3$ from different terms of the Kohn-Sham functional (1) are presented. It
is seen that the restoring force for E1 GR is determined by the electron-ion
interaction (ei) only. With increasing L, the electron-electron contribution
(ee) raises and starts to compensate the (ei) Coulomb part. Simultaneously,
the kinetic energy term grows. For high L, the contribution of the total
Coulomb interaction goes to zero and all the restoring force is determined
by the kinetic energy. The purely volume exchange correlation term (xc)
which within the LDA depends only on the electron density does not
contribute to $m_3$.

The restoring force should not be confused with the residual interaction. As
is seen from Eq. \ref{eq:ELres}, the residual interaction, unlike the
restoring force, has only the (ee)- and (xc)-terms for any L (where the
(ee)-term dominates).

\section{Anharmonicity and Multiphonon GR}

How much harmonic are the GR in metal clusters? How strong is the mixing of
one and two phonon states? Investigations performed within different
approaches$^{71-73}$ give contradictory answers for one-phonon GR. While the
shell-model calculations 
found for E1 GR in $Na_{20}$ some signals of anharmonicity \cite{KML_94},
other studies predict the harmonic character for M2(spin-dipole) and EL GR 
\cite{Cat_93,CRS97}. It should be noted that these studies have been done
for rather small clusters with $N\le 20$. In this size region the GR energy
lies safely below the lowest 2p-2h configurations, what does not favor the
anharmonic effects. This picture can change in larger clusters where GR
approach the region of 2p-2h configurations.

The calculations \cite{Cat_93} predict a noticeable anharmonicity for most
of {\it double} (two-phonon) GR placed at 8-15 eV. These GR exhibit a weak
mixing with one-phonon states but a considerable fragmentation between
two-phonon configurations. Most strong effect is expected for some $0^{+}$
double GR, for example, for $(1^{-}\otimes 1^{-})_{0^{+}}$ in $Na_{21}^{+}$.
With the appearance of new experimental techniques allowing investigation of
multiple GR these predictions are quite important. The techniques use
non-intense femtosecond lasers \cite{S98} or exploit collisions of a cluster
with highly charged ions \cite{G97}. Quite recently the multiple GR
constructed from 3-4 dipole plasmons has been observed in $Na_{93}^{+}$ \cite
{S98}.

\section{Magnetic GR}

Like in atoms, the spin-orbital interaction in metal clusters is negligible
and thus spin and orbital collective magnetic modes are well decoupled. The
separation of these two modes in MC is easier than in nuclei.

\subsection{Spin-Multipole GR}

Magnetic multipole resonances (ML) of spin character caused by the external
field $Q_L=\sum_{j=1}^Nr_j^LY_{L0}\sigma _j^z$ 
were studied within the SRA and RPA \cite
{LiC_96spin,SL_97spin,SBBN_93spin,MCSRe_96spin}. 
For $L=1$ the operator $Q_1\sim \sum_{j=1}^Nz_j\sigma _j^z$ provides the
opposite shifts of spin-up and spin-down electrons in z-direction. Unlike
the electric resonances , the residual interaction for ML GR is defined only
by the exchange and correlations ((xc)-term) since only the (xc)-term
depends on the magnetization density (see Eq. \ref{eq:MLres}) . Therefore
the study of ML resonances can provide a valuable information about
(xc)-effects in clusters.

Approximating the electron density by 
the expression $n_0=n_{00}/(1+exp((r-R)/a)$, one gets the energy for ML GR 
\cite{SL_97spin} 
\begin{eqnarray}
\omega _{ML} &=&\sqrt{\frac{m_3}{m_1}}=\hbar [\frac 25(2L+1)(L-1)\frac{\beta
_F^2}{R^2}+\frac{e^2}{m}4\pi aL\frac{n_0}{R^{2L-1}}  \nonumber \\
&+&\frac 1mL(v_{xc}^{(02)}(n_0,m_0)-v_{xc}^{(20)}(n_0,m_0))\frac{n_0}{6aR}%
]^{1/2}  \label{eq:omML}
\end{eqnarray}
where $v_{xc}^{(pq)}(n_0,m_0)=\frac{d^p}{dn^p}\frac{d^q}{dm^q}%
v_{xc}^{(pq)}(n,m)\mid _{(n=n_0,m=m_0)}$. For other notation see Eq. \ref
{eq:omEL}. As compared to Eq. \ref{eq:omEL} for EL GR, Eq. (9), 
was derived taking into account the spill-out effect. Furthermore, due to
the presence of the spin in the operator $Q_L$, the (xc)-term now
contributes to $m_3$, unlike the case of EL GR. However, the exchange
contributions (Pauli principle) to $v_{xc}^{20}$ and $v_{xc}^{02}$ are the
same and then, only correlation effects enter Eq. \ref{eq:omML}. The
energies of spin ML resonances decrease with N and go to zero for large
sizes. The larger $L$, the higher the GR energy. The behavior of ML GR much
depends on the diffuseness parameter $a$. 

\begin{table}[t]
\caption{ Relative contributions to $m_3$ for $Na_{92}$ and $Na_{912}$ :
kinetic energy $(m_3(T))$, correlation $(m_3(c))$ and total Coulomb
interaction $(m_3(C))$. The data are extracted from the Fig. 1 of Ref. 
\protect\cite{SL_97spin}. }
\label{tab:m3ML}\vspace{0.4cm}
\par
\begin{center}
\begin{tabular}{|c|c|c|c|c|c|c|}
\hline
& \multicolumn{3}{|c|}{$Na_{92}$} & \multicolumn{3}{|c|}{$Na_{912}$} \\ 
\cline{2-7}
L & $m_3(T)$ & $m_3(c)$ & $m_3(C)$ & $m_3(T)$ & $m_3(c)$ & $m_3(C)$ \\ \hline
1 & 0 & 0.77 & 0.23 & 0 & 0.30 & 0.70 \\ 
2 & 0.50 & 0.49 & 0.01 & 0.31 & 0.68 & 0.01 \\ 
5 & 0.76 & 0.20 & 0.04 & 0.59 & 0.38 & 0.03 \\ \hline
\end{tabular}
\end{center}
\end{table}

Table \ref{tab:m3ML} demonstrates that the restoring force for
spin-multipole GR differs from the electrical GR case. Namely, the
contribution of correlations\cite{VWN80} dominates for $L=1$ and 2 and
remains to be considerable for larger $L$. 
The correlation term includes long-range RPA correlations$^{76-78}$, 
short-range correlations \cite{LST94} and others. The correlations greatly
influence both static and dynamical characteristics of MC$^{37,38,76-78}$
and their investigation is very important.

\subsection{Orbital GR}

Since the number of atoms in MC can be much more than the number of nucleons
in nuclei, much larger values of the single-particle orbital moment can be
achieved what can give origin to very strong orbital magnetic multipole
resonances, orbital ML GR. That is, clusters, as nuclei, can exhibit orbital
ML GR like,``scissors'', twist mode, etc., nevertheless with a much stronger
strength \cite{LiS89_ZPD,Ba94}. 

Investigations of the specific low-energy orbital M1 GR, which can exist
only in {\it deformed} clusters demonstrated that it can serve as a good
indicator of the cluster quadrupole deformation $^{7,81-83}$. 
Indeed, in some cases the deformation splitting of E1 GR is washed out by
other effects and is not enough distinctive to get a reliable information on
cluster deformation. Then the orbital M1 GR can be used for this aim.
Macroscopically, this resonance is treated as small-angle rigid rotations of
the ellipsoid of valence electrons against the ionic ellipsoid. Such
collective mode was shown to be coupled with the quadrupole component, $%
\bigtriangledown (yz)$, of the displacement field \cite{LiS89_ZPD,LS89}. The
orbital M1 GR has the counterpart in deformed nuclei, well known as the
''scissors'' mode\cite{IP78}. The latter describes the rotations of the
neutron ellipsoid against the proton one. The orbital M1 GR is represented
by $K^\pi =1^{+}$ states ($K$ is the angular-momentum projection) with a low
excitation energy and strong M1 transitions to the ground state. For Na
clusters these characteristics are estimated 
as \cite{LiS89_ZPD,LS89} $\omega _{M1}=4.6\beta _2N_e^{-1/3}(1+5\frac{\omega
_0}{\omega _p})^{-1/2}$ eV and $B(M1)=1.1\beta _2N_e^{4/3}\mu _b^2$ where $%
\beta _2$ is the deformation parameter, $B(M1)$ is the reduced transition
probability and $\omega _0$ is the harmonic oscillator frequency. Both $%
\omega _{M1}$ and $B(M1)$ are proportional to the deformation parameter and
so the orbital M1 GR survives only in deformed clusters.

\begin{table}[t]
\caption{The excitation energy and strength (within the interval 0-1 eV) of
orbital M1 GR, calculated within the SRPA \protect\cite{NKS_Nas,NKSI_SN}.
See the text for notation. }
\label{tab:OM1}\vspace{0.4cm}
\par
\begin{center}
\begin{tabular}{|c|c|c|c|c|c|}
\hline
& $Na^+_{15}$ & $Na^+_{27}$ & $Na^+_{35}$ & $Na^+_{119}$ & $Na^+_{295}$ \\ 
\hline
$\beta_2$ & 0.32 & 0.23 & -0.23 & 0.25 & 0.24 \\ 
$\omega _{M1}$, eV & 0.63 & 0.29 & 0.35 & 0.26 & 0.21 \\ 
$B(M1), \mu_b^2$ & 27 & 56 & 41 & 229 & 757 \\ \hline
\end{tabular}
\end{center}
\end{table}

The results of the first realistic RPA calculations for orbital M1 GR \cite
{NKS_Nas,NKSI_SN} are given in Table 3. It is seen that this resonance has
low excitation energies. The most remarkable result is that already in
clusters with about 300 atoms, the orbital M1 GR strength reaches very high
values, 700-800 $\mu _b^2$. This GR is described in detail in Ref.\cite
{NKSI_SN} of the present Proceedings.

\section{Other GR in Atomic Clusters}

As compared to nuclei, atomic clusters provide many specific manifestations
of E1 GR. For example, clusters embedded in a dielectric matrix demonstrate
a strong screening effect: the matrix screens the residual interaction
between valence electrons in a cluster, which results in the considerable
decrease of E1-energy \cite{RS93}. In mixed and coated clusters the impurity
(or coated) atoms much influence both the ground state and properties of E1
GR (see, e.g. Refs.$^{86-88}$). 
In the fullerene $C_{60}$, two E1 GR are known as determined by weakly
bonded $\pi $ electrons and strongly bonded $\sigma $ electrons (see, e.g.
Ref. \cite{GL93_C60}).

$^3He$ and $^4He$ clusters representing collections of fermions ($^3He$
atoms) and bosons ($^4He$ atoms), respectively, should be mentioned. In $%
^3He $ clusters just $^3He$ atoms (not valence electrons) form a mean field
with quantum shells \cite{B90,S91,WR93}. These clusters are characterized by
strong surface effects. Unlike nuclei and MC, $^3He$ clusters represent the
case of {\it one-component} Fermi-system and, so, have no E1 GR. At the same
time, the study of other EL GR reveals new possibilities, for instance, the
comparison of the GR properties in Fermi ($^3He$ clusters) \cite{WR92,S91_He}
and Bose ($^4He$ clusters)\cite{CS90} systems.

\section*{Summary}

Giant resonances in atomic clusters have been observed. Being much similar
to their counterparts in atomic nuclei, GR in MC demonstrate, at the same
time, numerous exciting peculiarities. 
The unique situation takes place now in many-body physics where, in addition
to atoms and atomic nuclei, a new family of small Fermi systems (MC,
fullerenes, $He^3$ clusters, quantum dots) appears. This greatly enlarges
our possibilities in many-body studies. 
All mentioned systems possess, in a different extent, a mean field with
quantum shells. 

It should be noted that atomic clusters are attractive both for fundamental
studies and practical applications \cite{Ne_JINR}. Last achievements
(creation of new materials, machinning superhard surfaces, creation of
extremely large energy densities in a matter, catalysis, microelectronics,
microcomputering, etc.) show that, due to atomic clusters, one may expect in
a recent future a remarkable progress in many high-tech fields.

\section{Acknowledgments}

We are grateful to M. Schmidt and H. Haberland for communication the
experimental results and to the Organizing Committee of the Workshop for the
financial support of the attendance. The work was also partly supported by
CAPES (V.O.N.) and FINEP Brasil (V.O.N. and F.F.S.C.).

\newpage
{\bf FIGURE CAPTIONS}.

{\bf Figure 1}. E1 GR (dipole plasmon) in $K^+_{21}$ and $Li^+_{21}$ calculated in
the framework of the SRPA with (down) and without (up) the nonlocal ICE
contribution \cite{KNRS_EPJ98}. For $Li^+_{21}$ the photoabsorption
experimental data \cite{EH98} ($\triangle$) in \AA$^2/N_e$ are
compared.

{\bf Figure 2}. {\it Top:} E1 GR in spherical Na clusters from different size
regions. The SRPA results \cite{KNRS_EPJ98} are shown as bars for
every RPA state to demonstrate the Landau damping and as smoothed by a
Lorentz function (of the width 0.25 eV) to simulate the typical thermal
broadening of the plasmon. The length of the bars is rescaled by the factor
1/2.55 to fit the scale of the smoothed strength. The photoabsorption
experimental data are taken from Ref. \cite{Hab_PRL}. {\it Bottom:}
The number of dipole particle-hole configurations, as a function of the
energy, corresponding to $\Delta {\cal N}$=1 (light dotted bricks), $\Delta 
{\cal N}$=3 (dashed bricks), $\Delta {\cal N}$=5 (dark dotted bricks) and $%
\Delta {\cal N}\geq$7 (unfilled bricks) dipole transitions in $Na$ clusters
presented in the top of the figure. ${\cal N}$ is the principal shell
quantum number. The arrows mark centroid energies of the plasmon. 

{\bf Figure 3.} E1 GR in deformed Na clusters. The SRPA results \cite
{NK_Tsu} (curves and bars) are given by the same way as in Figure 2. 
The experimental data are taken from Ref.\cite{SH97}.
The deformation parameter $\beta_2$ is extracted from the experiment 
\cite{SH97} following the prescription\cite{LiS89_ZPD}. 

{\bf Figure 4.} E2 and E3 GR in $Na^+_{59}$ calculated within the SRPA. 

\section*{References}

\begin{thebibliography}{99}
\bibitem{}  

\bibitem{Ne92}  
V.O. Nesterenko, {\em Sov. J. Part. Nucl.} {\bf 23}, 1665 (1992).

\bibitem{He93}  
W.A. de Heer, {\em Rev. Mod. Phys.} {\bf 65}, 611 (1993).

\bibitem{Br93}  
M. Brack, {\em Rev. Mod. Phys.} {\bf 65}, 677 (1993).

\bibitem{Bre_94rev}  
C. Brechignac and J.P. Connerade, {{\em J. Phys.} B} {\bf 27}, 3795 (1994).

\bibitem{Ne_Dub}  
V.O. Nesterenko, W. Kleinig and V.V. Gudkov, in Proc. Intern. Conf. {\em %
Nuclear Structure and Related topics}, ed. S.N. Ershov, R.V. Jolos and V.V.
Voronov (JINR, Dubna, 1997) p. 322. 

\bibitem{Se_SRA_PRB89}  
Ll. Serra {\em et al}, {{\em Phys. Rev.} B} {\bf 39}, 8247 (1989).

\bibitem{LiS89_ZPD}  
E. Lipparini and S. Stringari, {{\em Z. Phys.} D} {\bf 18}, 193 (1991).

\bibitem{LiC_96spin}  
E. Lipparini and M. Califano, {{\em Z. Phys.} D} {\bf 37}, 365 (1996).

\bibitem{SL_97spin}  
Ll. Serra and E. Lipparini, {{\em Z. Phys.} D} {\bf 42}, 227 (1997). 

\bibitem{RB90}  
P.-G. Reinhard and M. Brack, {{\em Phys. Rev.} A} {\bf 41}, 5568 (1990).

\bibitem{ReGB_AP96}  
P.-G. Reinhard, O. Genzken and M. Brack, {\em Ann. Phys. (Leipzig)} {\bf 5},
576 (1996). 

\bibitem{GJ92}  
C. Guet and W.R. Johnson, {{\em Phys. Rev.} B} {\bf 45}, 11283 (1992). 

\bibitem{Eka84JM}  
W. Ekardt,{\em Phys. Rev. Lett.} {\bf 52}, 1925 (1984).

\bibitem{Be84JM}  
D.E. Beck, {{\em Phys. Rev.} B} {\bf 30}, 6935 (1984).

\bibitem{PE92}  
J.M. Pacheco and W. Ekardt, {\em Ann. Phys. (Leipzig)} {\bf 1}, 254 (1992).

\bibitem{YanB_PR91}  
C. Yannouleas and R.A. Broglia, {{\em Phys. Rev.} A} {\bf 44}, 5793 (1991).

\bibitem{Yan_CPL}  
C. Yannouleas, {\em Chem. Phys. Lett.} {\bf 193}, 587 (1992).

\bibitem{YVB93}  
C. Yannouleas, E. Vigezzi and R.A. Broglia, {{\em Phys. Rev.} B} {\bf 47},
9849 (1993). 

\bibitem{SBJL93}  
Ll. Serra, G.B. Bachelet, N. Van Giai and E. Lipparini, {{\em Phys. Rev.} B} 
{\bf 48}, 14708 (1993).

\bibitem{ASB95}  
F. Alasia {\em et al}, {{\em Phys. Rev.} B} {\bf 52}, 8488 (1995).

\bibitem{SLG95}  
Ll. Serra, E. Lipparini and N. Van Giai, {\em Europhys. Lett.} {\bf 29}, 445
(1995).

\bibitem{BG95}  
S.A. Blundell and C. Guet,{{\em Z. Phys.} D} {\bf 33}, 153 (1995).

\bibitem{CCG95}  
F. Catara, Ph. Chomaz and N. Van Giai, {{\em Z. Phys.} D} {\bf 33}, 219
(1995).

\bibitem{MRM94}  
B. Montag, P.-G. Reinhard and J. Meyer, {{\em Z. Phys.} D} {\bf 32}, 125
(1994).

\bibitem{YB96}  
K. Yabana and G.F. Bertsch, {{\em Phys. Rev.} B} {\bf 54}, 4484 (1996).

\bibitem{Rubrev}  
A. Rubio, J.A. Alonso, X. Blase, S.G. Louie, to appear in {{\em Int. J. Mod.
Phys.} B} , 1998 (. ) 

\bibitem{NK_PS}  
V.O. Nesterenko and W. Kleinig, {\em Phys. Scr.} {\bf T56}, 284 (1995).

\bibitem{Ne_ZPD}  
V.O. Nesterenko, W. Kleinig and V.V. Gudkov, {{\em Z. Phys.} D} {\bf 34},
271 (1995).

\bibitem{Ne_PRC}  
V.O. Nesterenko, W. Kleinig, V.V. Gudkov and J. Kvasil,, {{\em Phys. Rev.} C}
{\bf 53}, 1632 (1996).

\bibitem{Ne_PRA}  
V.O. Nesterenko, W. Kleinig, V.V. Gudkov, N. Lo Iudice and J. Kvasil, {{\em %
Phys. Rev.} A} {\bf 56}, 607 (1997).

\bibitem{NK_Tsu}  
V.O. Nesterenko and W. Kleinig, in Proc. Intern. Symp. {\em Similarities and
Differences between Atomic Nuclei and Clusters} (Tsukuba, Japan, 1997), ed.
Y. Abe, I. Arai, S.M. Lee and K. Yanaba, AIP Conference Proceedings 416,
(Woodbury, New York, 1998) p.77.

\bibitem{NKK_Pra}  
V.O. Nesterenko, W. Kleinig and J.Kvasil, in Proc. Intern. Conf. {\em Atomic
Nuclei and Metallic Clusters} (Prague, Czech Republic, 1997), ed. P.Alexa, 
{\em Czech. J. Phys.} {\bf 48}, 745 (1998).

\bibitem{KNRS_EPJ98}  
W. Kleinig, V.O. Nesterenko, P.-G. Reinhard, Ll. Serra, to be published in 
{\em Eur. Phys. J D}; Preprint JINR, E17-98-126, Dubna, 1998.

\bibitem{R_SRPA}  
J. Babst and P.-G. Reinhard, {{\em Z. Phys.} D} {\bf 42}, 209 (1997).

\bibitem{YB_APNY91}  
C. Yannouleas and R.A. Broglia, {\em Ann. Phys. (N.Y.)} {\bf 217}, 105
(1991). 

\bibitem{KS65}  
W. Kohn and L. J. Sham, {\em Phys. Rev. } {\bf 140}, A1133 (1965).

\bibitem{GL76}  
O. Gunnarsson and B.I. Lundqvist, {{\em Phys. Rev.} B} {\bf 13}, 4274 (1976).

\bibitem{VWN80}  
S.H. Vosko, L. Wilk and M. Nusair, {\em Can. J. Phys.} {\bf 58}, 1200
(1980). 

\bibitem{C85}  
K. Clemenger,{{\em Phys. Rev.} B} {\bf 32}, 1359 (1985).

\bibitem{RFB95}  
S.M. Reinmann, S. Frauendorf and M. Brack, {{\em Z. Phys.} D} {\bf 34}, 125
(1995).

\bibitem{NHM90}  
H. Nishioka, K.I. Hansen and B.R. Mottelson, {{\em Phys. Rev.} B} {\bf 42},
9377 (1990).

\bibitem{FP96}  
S. Frauendorf and V.V. Pashkevich, {\em Ann. Phys. (Leipzig)} {\bf 5}, 34
(1996). 

\bibitem{Hab_PRL}  
Th. Reiners {\it et al}, {\em Phys. Rev. Lett.} {\bf 74}, 1558 (1995).

\bibitem{Bre91}  
C. Brechignac {\it et al}, {{\em Z. Phys.} D} {\bf 19}, 1 (1991).

\bibitem{Sel89}  
K. Selby {\it et al}, {{\em Phys. Rev.} B} {\bf 40}, 5417 (1989).

\bibitem{Borg93}  
J. Borgreen {\it et al}, {{\em Phys. Rev.} B} {\bf 48}, 17507 (1993).

\bibitem{Na10_PRL}  
Ch. Ellert {\it et al}, {\em Phys. Rev. Lett.} {\bf 75}, 1731 (1995).

\bibitem{Ti_92GR}  
J. Tiggesbaumker {\it et al}, {\em Chem. Phys. Lett.} {\bf 190}, 42 (1992).

\bibitem{Me97}  
P. Meibom {\it et al}, {{\em Z. Phys.} D} {\bf 40}, 258 (1997). 

\bibitem{SBBN_93spin}  
Ll. Serra {\it et al}, {{\em Phys. Rev.} A} {\bf 47}, R1601 (1993).

\bibitem{MCSRe_96spin}  
L. Mornas {\it et al}, 
{{\em Z. Phys.} D} {\bf 38}, 73 (1996). 

\bibitem{Mie}  
G. Mie, {\em Ann. Phys. (N.Y.)} {\bf 25}, 377 (1908).

\bibitem{Ar_NC89}  
S. Arvati {\it et al},{{\em Nuovo Cimento} D} {\bf 7}, 1063 (1989). 

\bibitem{BHS82}  
G.B. Bachelet, D.R. Hamman and M. Schl$\ddot u$ter, {{\em Phys. Rev.} B} 
{\bf 26}, 4199 (1982).

\bibitem{BCC89}  
G.B. Bachelet, D.M. Ceperley and M.G.B. Chiochetti, {\em Phys. Rev. Lett.} 
{\bf 62}, 2088 (1989).

\bibitem{L96}  
J. Lerm$\acute e$, {{\em Phys. Rev.} B} {\bf 54}, 14158 (1996). 

\bibitem{EH98}  
C. Ellert and H. Haberland, private communication.

\bibitem{SR97}  
Ll. Serra and A. Rubio, {\em Phys. Rev. Lett.} {\bf 78}, 1428 (1997). 

\bibitem{B90}  
S. Bjornholm, {\em Contemp. Phys.} {\bf 31}, 309 (1990).

\bibitem{MonRei_temp}  
B.Montag and P.-G.Reinhard, {{\em Phys. Rev.} B} {\bf 51}, 14686 (1995).

\bibitem{BL95}  
A. Bulgac and C. Lewenkopf, {\em Europhys. Lett.} {\bf 31}, 519 (1995).

\bibitem{HB67}  
{\em Hand Book of Chemistry and Physics}  (Chemical Rubber, Cleveland,
1967), p. 56.

\bibitem{BK_Pra}  
V. Bona$\hat c$i$\acute c$-Kouteck$\acute y$, {\it et al},  in Proc. Intern.
Conf. {\em Atomic Nuclei and Metallic Clusters}  (Prague, Czech Republic,
1997), ed. P.Alexa, {\em Czech. J. Phys.} {\bf 48}, 637 (1998). 

\bibitem{KLM_95UJM}  
M. Koskinen, P.O. Lipas and M. Manninen, {\em Europhys. Lett.} {\bf 30}, 519
(1995).

\bibitem{FP96Er}  
S. Frauendorf and V.V. Pashkevich, In Proc. Int. School  "Large Clusters of
Atoms and Molecules" (Erice, 1996), ed.  T.P.Martin, (1996) 201.

\bibitem{MHMRB}  
B. Montag, {\it et al}, 
{{\em Phys. Rev.} B} {\bf 52}, 4775 (1995).

\bibitem{Hi_ZPD}  
Th. Hirschmann, B. Montag and J. Mejer, {{\em Z. Phys.} D} {\bf 37}, 63
(1996).

\bibitem{YL97}  
C. Yannouleas and U. Landman, {{\em Phys. Rev.} B} {\bf 51}, 1902 (1997). 

\bibitem{SH97}  
M. Schmidt and H. Haberland, private communication. 

\bibitem{NS80}  
J.R. Nix and A.J. Sierk, {{\em Phys. Rev.} C} {\bf 21}, 396 (1980). 

\bibitem{KML_94}  
M. Koskinen, M. Manien and P.O. Lipas, {{\em Phys. Rev.} B} {\bf 49}, 8418
(1994).

\bibitem{Cat_93}  
F. Catara, Ph. Chomaz and N. Van Giai, {{\em Phys. Rev.} B} {\bf 48}, 18207
(1993).

\bibitem{CRS97}  
F. Calvayrac, P.-G. Reinhard and E. Suraud, {\em Ann. Phys. (N.Y.)} {\bf 255}%
, 125 (1997).

\bibitem{S98}  
R. Schplipper {\it et al}, {\em Phys. Rev. Lett.} {\bf 80}, 1194 (1998).

\bibitem{G97}  
C. Guet {\it et al}, {{\em Z. Phys.} D} {\bf 40}, 317 (1997). 

\bibitem{R92}  
P.-G. Reinhard, {{\em Phys. Lett.} A} {\bf 169}, 281 (1992).

\bibitem{YCV95}  
C. Yannouleas, F. Catara and N. Van Giai, {{\em Phys. Rev.} B} {\bf 51},
4569 (1995).

\bibitem{CPSV96}  
F. Catara, G. Piccitto, M. Sanbataro and N. Van Giai, {{\em Phys. Rev.} B} 
{\bf 54}, 17536 (1996).

\bibitem{LST94}  
E. Lipparini, Ll. Serra and K. Takayanagi, {{\em Phys. Rev.} B} {\bf 49},
16733 (1994). 

\bibitem{Ba94}  
S.I. Bastrukov, {\em J. Moscow Phys. Soc.} {\bf 4}, 57 (1994).

\bibitem{LS89}  
E. Lipparini and S. Stringari, {\em Phys. Rev. Lett.} {\bf 63}, 570 (1989).

\bibitem{NKS_Nas}  
V.O. Nesterenko, W. Kleinig and F.F. de Souza Cruz,  to be published in
Proc. XXII Intern. Workshop on  Condensed Matter Theories (Nashville, US,
1998).

\bibitem{NKSI_SN}  
V.O. Nesterenko, W. Kleinig, F.F. de Souza Cruz and N. Lo Iudice,  in Proc.
of Intern. Workshop {\em Collective excitations in Fermi  and Bose Systems}
(Serra Negra, Brazil, 1998).

\bibitem{IP78}  
N. Lo Iudice and F. Palumbo, {\em Phys. Rev. Lett.} {\bf 41}, 1532 (1978). 

\bibitem{RS93}  
A. Rubio and Ll. Serra, {{\em Phys. Rev.} B} {\bf 48}, 18222 (1993).

\bibitem{YJK92}  
C. Yannouleas, P. Jena and S.N. Khanna, {{\em Phys. Rev.} B} {\bf 46}, 9751
(1992).

\bibitem{A94}  
J.A. Alonso, {\em Phys. Scr.} {\bf T55}, 177 (1994).

\bibitem{RALS94}  
A. Rubio, J.A. Alonso, J.M. Lopez and M.J. Spott, {{\em Phys. Rev.} B} {\bf %
49}, 17397 (1994).

\bibitem{GL93_C60}  
N. Van Giai and E. Lipparini, {{\em Z. Phys.} D} {\bf 27}, 193 (1993). 

\bibitem{S91}  
S. Stringari, {{\em Z. Phys.} D} {\bf 20}, 219 (1991).

\bibitem{WR93}  
S. Weisgerber and P.-G. Reinhard, {\em Ann. Phys. (Leipzig)} {\bf 2}, 666
(1993).

\bibitem{WR92}  
S. Weisgerber and P.-G. Reinhard, {{\em Z. Phys.} D} {\bf 23}, 275 (1992).

\bibitem{S91_He}  
Ll. Serra {\it et al}, {\em Phys. Rev. Lett.} {\bf 67}, 2311 (1991).

\bibitem{CS90}  
M. Cassas and S. Stringari, {\em J. Low. Temp. Phys.} {\bf 79}, 135 (1990). 

\bibitem{Ne_JINR}  
V.O. Nesterenko, {\em JINR News}, 1/1998, ISSN 0134-4811, Dubna, p. 6.
\end{thebibliography}
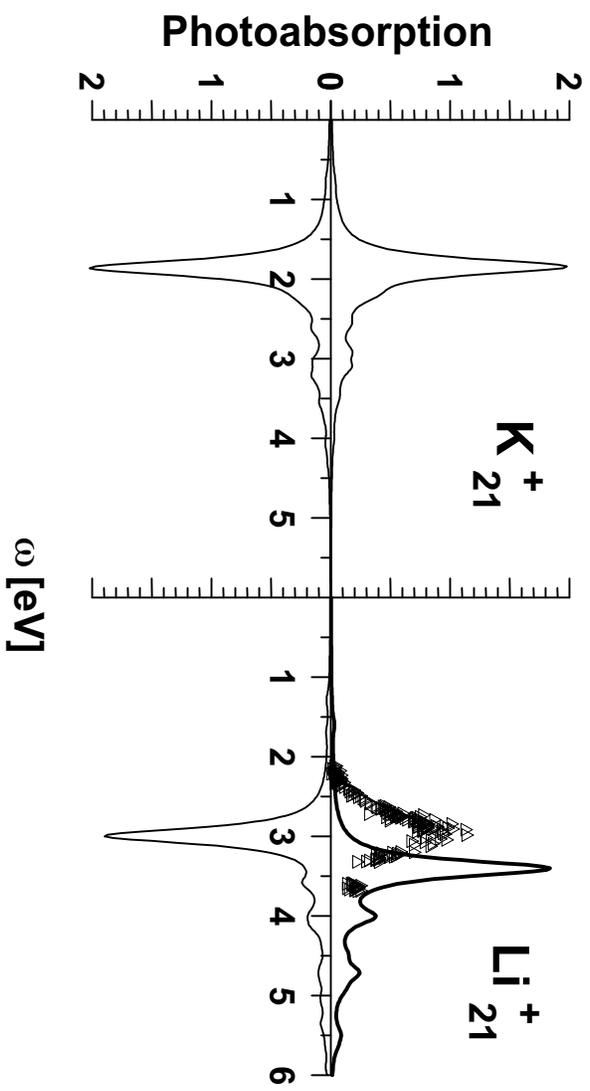
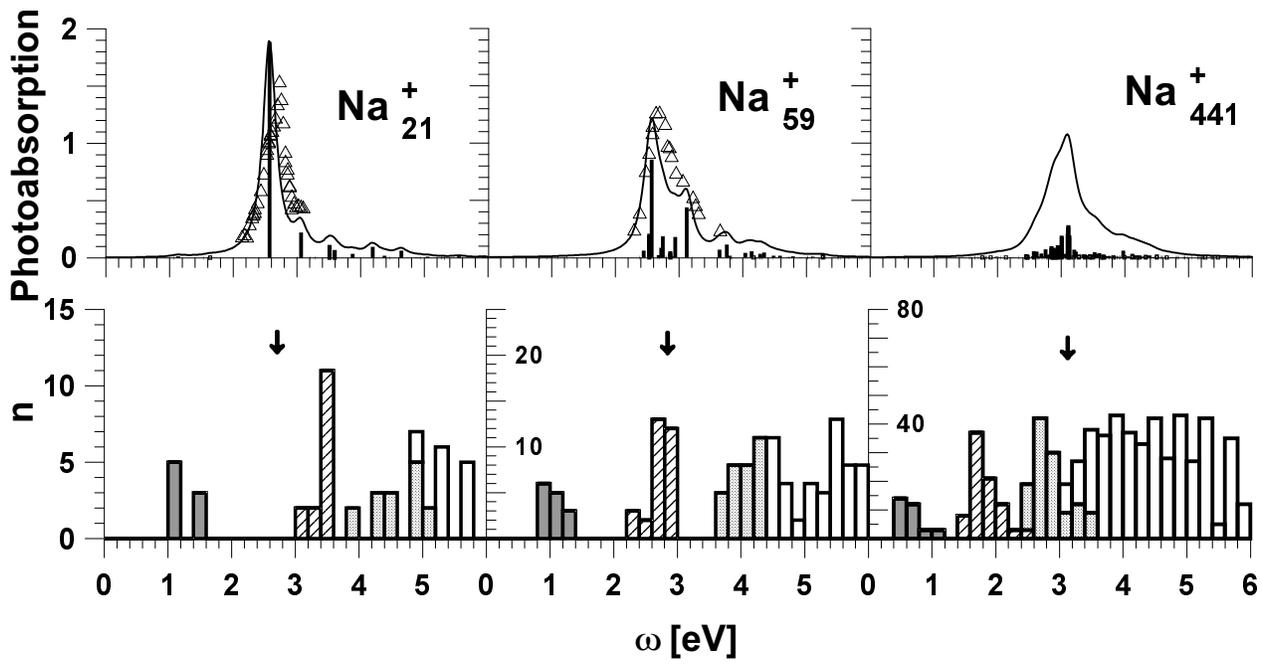
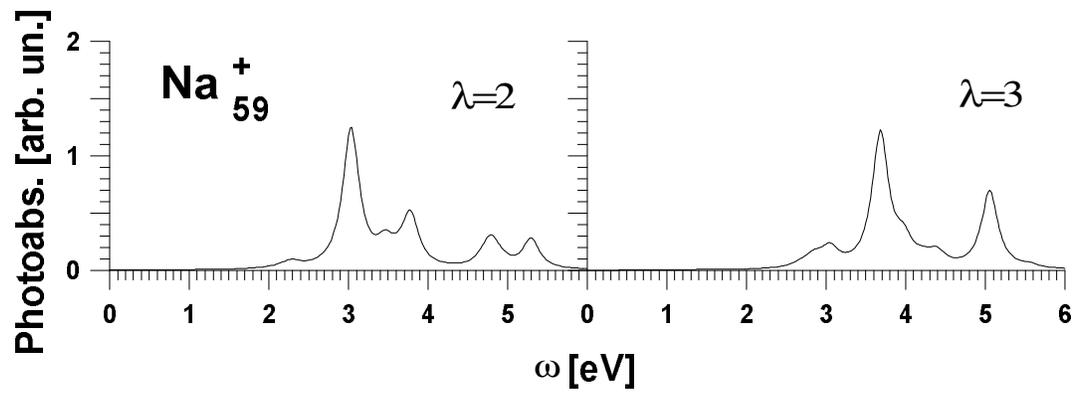
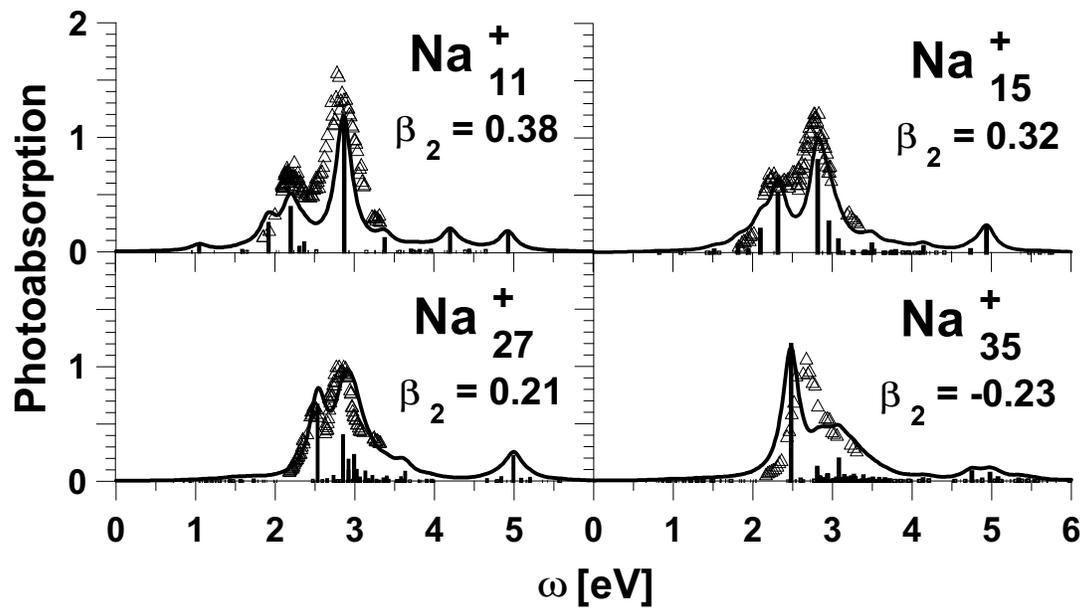
\end{document}